# Near to Mid-Infrared Spectroscopy of (65803) Didymos as observed by JWST: Characterization Observations Supporting the Double Asteroid Redirection Test

Andrew S. Rivkin[1], Cristina A. Thomas[2], Ian Wong[3], Benjamin Rozitis[4], Julia de León[5], Bryan Holler[6], Stefanie N. Milam[3], Ellen S. Howell[7], Heidi B. Hammel[8], Anicia Arredondo[9], John R. Brucato[10], Elena M. Epifani[11], Simone Ieva[11], Fiorangela La Forgia[12], Michael P. Lucas[13], Alice Lucchetti[14], Maurizio Pajola[14], Giovanni Poggiali[10], Jessica N. Sunshine[15], and Josep M. Trigo-Rodríguez[16]

1. Johns Hopkins University Applied Physics Laboratory, 11100 Johns Hopkins Road, Laurel, MD, 20723, USA; andy.rivkin@jhuapl.edu
2. Northern Arizona University, Department of Astronomy and Planetary Science, P.O. Box 6010, Flagstaff, AZ 86011, USA
3. NASA Goddard Space Flight Center, 8800 Greenbelt Road, Greenbelt, MD 20771, USA
4. School of Physical Sciences, The Open University, Milton Keynes, UK
5. Instituto de Astrofísica de Canarias (IAC), Santa Cruz de Tenerife, Spain
6. Space Telescope Science Institute, 3700 San Martin Drive, Baltimore, MD 21218, USA
7. Lunar & Planetary Laboratory, University of Arizona, Tucson, AZ 85721, USA
8. Association of Universities for Research in Astronomy, 1212 New York Avenue NW, Suite 450, Washington, DC 20005, USA
9. Southwest Research Institute, San Antonio TX, USA
10. INAF- Osservatorio Astrofisico di Arcetri, Firenze, Italy
11. INAF - Osservatorio Astronomico di Roma, Via Frascati 33 - 00078 Monte Porzio Catone (RM), Italy
12. Department of Physics and Astronomy 'G. Galilei', University of Padova, Vicolo dell'Osservatorio 3, I-35122 Padova, Italy
13. Department of Civil & Environmental Engineering & Earth Sciences, University of Notre Dame, IN
14. INAF- Osservatorio Astronomico di Padova, Padova, Italy
15. Departments of Astronomy and Geology, University of Maryland, College Park, MD 20742-2421, USA
16. Institute of Space Sciences (CSIC-IEEC), Campus UAB, Carrer de Can Magrans, s/n 08193 Cerdanyola del Vallés, Catalonia, Spain.

## 1. Abstract

The Didymos binary asteroid was the target of the Double Asteroid Redirection Test (DART) mission, which intentionally impacted Dimorphos, the smaller member of the binary system. We used the Near-Infrared Spectrograph (NIRSpec) and Mid-Infrared Instrument (MIRI) instruments on JWST to measure the 0.6–5 µm and 5–20 µm spectra of Didymos approximately two months after the DART impact. These observations confirm that Didymos belongs to the S asteroid class and is most consistent with LL chondrite composition as was previously determined from its 0.6—2.5-µm reflectance spectrum. Measurements at wavelengths > 2.5 µm show Didymos to have thermal properties typical for an S-complex asteroid of its size and to be lacking absorptions deeper than ~2% due to OH or $H_2O$. Didymos' mid-infrared emissivity spectrum is within the range of what has been observed on S-complex asteroids observed with Spitzer Space Telescope and is most consistent with emission from small (< 25 µm) surface particles. We conclude that the observed reflectance and physical properties make the Didymos system a good proxy for the type of ordinary chondrite asteroids that cross near-Earth space, and a good representative of likely future impactors.

## 2. Background

Didymos is a binary asteroid system named for its largest member, (65803) Didymos (diameter 761 ± 26 m; Daly et al. 2023). It also contains a satellite asteroid, Dimorphos (diameter 151 ± 5 m: Daly et al. 2023), famous for being the target of the NASA Double Asteroid Redirection Test (DART) mission (Rivkin et al., 2021). The DART spacecraft intentionally collided with Dimorphos as a test of the kinetic impactor planetary defense technique, and knowledge of its target's physical and compositional properties is an important component for interpreting the results of the impact experiment (Daly et al. 2023). DART's payload was a single, monochromatic camera (the Didymos Reconnaissance and Asteroid Camera for Optical navigation, or DRACO), and DART was accompanied by the Light Italian CubeSat for Imaging of Asteroids (LICIACube; Dotto et al. 2021), contributed by the Italian Space Agency. LICIACube carried two cameras, the monochromatic Liciacube Explorer Imaging for Asteroid (LEIA) and the Liciacube Unit Key Explorer (LUKE). LUKE was equipped with an RGB Bayer pattern filter, which provided the only spectral information from within the Didymos system. Until the arrival of the Hera mission (Section 7) in December 2026, spectral studies of the Didymos system must be done remotely.

Dimorphos orbits too close to Didymos to be separately resolvable by Earth-based or space-based optical telescopes, and therefore its discovery was made via radar detection and lightcurve studies (Pravec et al. 2003). At the time of the observations reported here, the centers of Didymos and Dimorphos were never separated by more than 0.1" from each other, even outside of mutual events. Didymos is roughly five times larger in diameter than Dimorphos (Daly et al. 2023) and therefore it has roughly 25 times the cross-sectional area as Dimorphos. This size difference between the components means that roughly 96% of flux from the system typically comes from Didymos.

Despite these challenges, several lines of evidence suggest that Dimorphos has a similar composition as Didymos (Pajola et al. 2022), and that we can reasonably estimate the composition of Dimorphos specifically from measurements of the Didymos-dominated flux. Currently there are no separate spectral measurements of near-Earth object (NEO) binary components in the literature but observations of unbound asteroid dynamical pairs, thought to have formed in a similar process as asteroid satellites (Vokrouhlický and Nesvorný 2008, Pravec et al. 2010) show similar spectral properties to each other (Moskovitz 2012, Moskovitz et al. 2019) that point to similar compositions. There is also evidence from the Didymos system itself: observations in the 0.34–0.81-µm range were made inside and outside of mutual events by Ieva et al. (2022), who found the spectra to be "substantially similar". Observations made shortly after the DART impact were dominated by ejecta derived from Dimorphos, which showed similar shapes in reflectance and polarization spectra as pre-impact data (Opitom et al. 2023, Bagnulo et al. 2023, Polishook et al. in review). Observations by Lazzarin et al. (2023) show the 0.5–0.9-µm spectrum of the Didymos system exhibiting an S-type spectrum through the impact period, accompanied by variation in spectral slope in October 2022 when the ejecta contribution to the flux was significant. These findings are all consistent with the current paradigms of the formation process for binary systems like Didymos (Margot et al. 2015; Lindsay et al., 2015), in which Dimorphos is derived from Didymos itself via rotational fission (Pajola et al. 2022). Consequently, measurements of Didymos provide our best current insight into the composition of Dimorphos until the arrival of the ESA Hera mission.

Beyond the need to study Didymos to support the DART mission, Didymos also serves as a representative near-Earth asteroid. While it was first assigned to the Xk spectral class based on a restricted wavelength range (Binzel et al. 2004), later measurements covering the entire 0.5—2.5-µm range showed Didymos to have a relatively typical S-type spectrum (de León et al. 2006), and Dunn et al. (2013) interpreted its spectrum as showing a composition similar to L or LL ordinary chondrites. It is generally accepted that the near-Earth asteroid population originated in the main asteroid belt, and large asteroid families are generally thought to be the most likely place for NEOs to originate. Richardson et al. (2022) suggested the Flora or Baptistina families as possible origin locations for Didymos, and the former has been associated with L/LL chondrites (Vernazza et al. 2008). Richardson et al. (2022) also noted that according to the Granvik et al. (2018) model Didymos has a > 80% likelihood of having arrived into near-Earth space via the $\nu_6$ secular precessional resonance. Binzel et al. (2019) notes the $\nu_6$ resonance is the most likely entry point to near-Earth space for asteroids with LL-chondrite-like spectra, while L-chondrite-like asteroids have no preferred source region when accounting for the uncertainties of the calculations.

We have obtained spectroscopic measurements of the Didymos system as part of Program 1245 using two instruments on JWST: NIRSpec (Jakobsen et al. 2022, Böker et al. 2023) and MIRI (Wells et al 2015, Wright et al. 2023). Below, we discuss the observations, the physical and compositional properties we can derive from them, and their implications.

## 3. Observations

Table 1 shows the observational circumstances for the JWST data. The JWST observations were centered on the low northern latitudes of Didymos, which passed through its northern spring equinox on 11 November 2023. For comparison, the sub-Earth latitude was -67° at the time of the DART impact on 26 September 2022 and remained southward of -30° until 21 October 2022.

NIRSpec observed Didymos on 28 November 2022 in fixed-slit mode with the PRISM grating, obtaining two exposures with an effective combined integration time of 112 seconds. The spectra cover 0.8—5.1 µm with a spectral resolving power of $\lambda/\Delta\lambda$ ~100. The target was dithered along the S200A1 slit between the two exposures, and the NRSRAPID readout pattern was used. MIRI measurements using the Medium Resolution Spectrometer (MRS) were made on 4 December 2022. The MIRI MRS consists of four integral field units (IFU), each dedicated to a particular wavelength range (Wells et al. 2015): Channel 1 (4.9—7.7 µm), Channel 2 (7.5—11.7 µm), Channel 3 (11.6—18.0 µm), and Channel 4 (17.7—27.9 µm). To sample the full wavelength range of each channel, three grating settings (or "sub-bands") are needed: short, medium, and long. The spectral resolving power varies from 1330 at the longest wavelengths to 3710 at the shortest wavelengths. Our Didymos observations utilized all three sub-bands in sequence, with each observation sampling all four channels simultaneously. A four-point dither pattern was used, and the total exposure time per sub-band was 677 seconds. In order to account for thermal emission from the telescope and the spatially varying contribution from zodiacal light, a separate dedicated MIRI observation of a nearby background field was executed immediately following the science observations of Didymos. The background observations were not dithered, with a total exposure time of 83 seconds per sub-band. The spectra from the separate channels and sub-bands were then combined to create a single spectral energy distribution (SED), which was the input for the thermal models and basis for the emissivity spectra discussed below.

According to the JPL Horizons ephemeris service (https://ssd.jpl.nasa.gov/horizons/app.html) the MIRI observations (Table 1) were obtained over roughly 40% of Didymos' rotation period. Given the nature of MIRI data collection, any spectral variation on Didymos' surface would potentially manifest as discontinuities between adjacent sub-bands, which are obtained with different grating settings (Section 4.2). As part of the reduction process, sub-bands were scaled to one another to provide a smooth spectrum where possible.

Figure 1 is a map showing the visible parts of Didymos observed in NIRSpec and MIRI medium sub-band observations. Shaded regions were both lit and in the JWST line of sight. The NIRSpec observations were centered very close to the opposite side of Didymos from where the MIRI long-wavelength sub-band observations were centered, but only 66° from where the short-wavelength sub-band observations were centered. The medium sub-band observations were observed between the long- and short-wavelength sub-bands (Table 1).

| Observation | Mid-time | Solar Phase angle | Solar Dist (AU) | JWST Dist (AU) | Sub-JWST lat/lon | Sub-Solar lat/lon | Mutual Event? |
|---|---|---|---|---|---|---|---|
| NIRSpec | 2022-11-28 00:58:23 | 54° | 1.084 | 0.159 | (230°, +21°) | (177°, +8°) | No |
| MIRI-long | 2022-12-04 07:09:03 | 45° | 1.109 | 0.171 | (31°, +25°) | (346°, +11°) | Dimorphos occulted by Didymos |
| MIRI-med | 2022-12-04 07:34:29 | | | | (97°, +25°) | (52°, +11°) | |
| MIRI-short | 2022-12-04 07:59:53 | | | | (164°, +25°) | (121°, +11°) | |

Table 1: Observing Circumstances for NIRSpec and MIRI Observations. The "mutual event" column describes whether Didymos and Dimorphos were both visible or if one object was occulting the other.

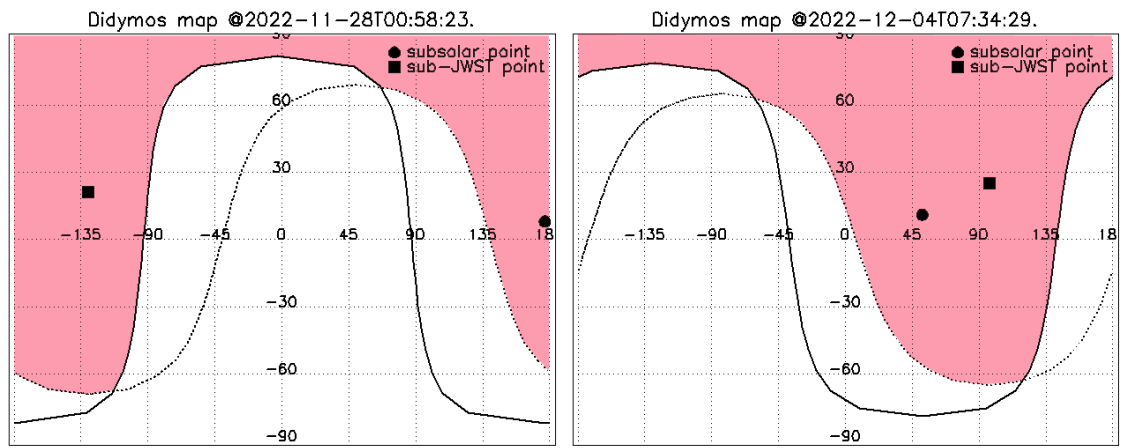

Figure 1: Maps of the surface coverage of Didymos from the NIRSpec and MIRI-Medium observations. Dotted lines separate the hemisphere visible from JWST from the side not visible from JWST, the solid line separates the lit hemisphere from the unlit hemisphere. Shaded regions show areas that were lit and in the line of sight from JWST. Note that these maps and the values in Table 1 do not take Didymos' shape into account, though Didymos is spheroidal and shape effects are small. The coordinate system used in this figure is defined by IAU standards (Archinal et al. 2018), with the prime meridian set to be the the zero longitude on 1 January 2000 at 12:00 UT.

While an ejecta tail composed of Dimorphos-derived material due to the DART impact was still evident in ground-based wide-field imagery of the system at the time of these observations (Moreno et al., 2023), photometric measurements show little to no contribution from the ejecta to the overall system brightness at those times (Lister et al., 2023). In addition, inspection of the MIRI frames shows Didymos' PSF to be consistent with what is expected for a point source (Figure 2). As a result, and given the relative sizes of Didymos and Dimorphos, we would expect the vast majority of flux from the system at the time of the JWST observations to come from Didymos rather than Dimorphos or Dimorphos-sourced material, making comparison to pre-impact observations appropriate. Additionally, the JPL Horizons ephemeris indicates that Dimorphos was being occulted by Didymos during the entire period of MIRI observations (Table 1). For comparison, the NIRSpec observations were obtained outside a mutual event and near maximum separation. As discussed in Section 1, we expect the results discussed below to apply to both Didymos and Dimorphos, given our understanding of asteroid satellite formation but we will specify cases where results may be Didymos-specific.

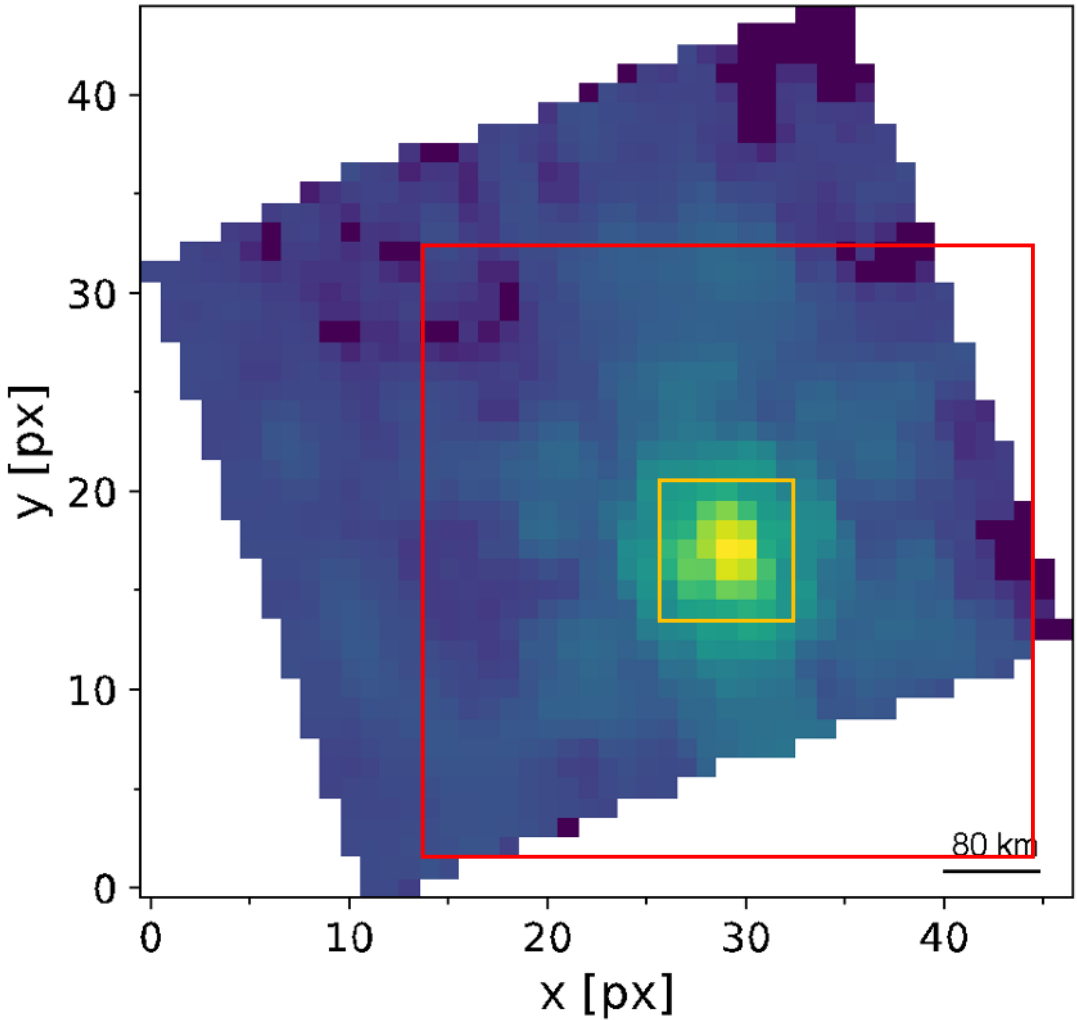

Figure 2: Median averaged slice through the MIRI MRS IFU showing Didymos. Note that Dimorphos was being occulted by Didymos during the entire period of MIRI observations. The measured median PSF FWHM is 0.332" in sub-band 2B, consistent with the predicted FWHM of 0.294". The measured FWHM across other sub-bands were found to be 0.9—1.2 times the predicted FWHM, broadly consistent with expectations for a point source. The yellow box marks the 7x7 pixel aperture used in our MIRI spectral extraction; the red box denotes the inner edge of the region used for background estimation.

## 4. Data Reduction

Data reduction and spectral extraction were carried out using a custom data processing framework, built around the official JWST pipeline (Bushouse et al. 2022), with a number of additional modifications. The analysis in this paper utilized Version 1.11.3 of the JWST pipeline and calibration reference files from Calibration Reference Data System (CRDS) build jwst_1100.pmap. Unless otherwise indicated, default settings were used for the various stages within the JWST pipeline.

4.1. NIRSpec Data Reduction

For the NIRSpec data, the uncalibrated Level 0 files were first passed through Stage 1 of the JWST pipeline, which converts the raw data, consisting of a series of regularly-spaced detector readouts in units of data number (DN), to Level 1 countrate images in units of DN/sec that are corrected for bias, dark current, and linearity effects. NIRSpec data are affected by correlated noise introduced by the detector readout process (Moseley et al. 2010), which manifests as vertical striping on the NIRSpec subarray images. We carried out a column-by-column readout noise correction of the Level 1 rate files by calculating the $3\sigma$-clipped median value from the top and bottom five pixels in each column (which are not illuminated) and subtracting it from the column flux array. We then passed the corrected rate files to Stage 2 of the JWST pipeline, which applies flat field and optical pathloss corrections and yields Level 2 distortion-corrected, wavelength-calibrated, and flux-calibrated data products.

We extracted the spectra from the Level 2 calibrated spectroscopic images (s2d files) using an empirical PSF fitting procedure. First, we masked out all pixels with a nonzero data quality flag, which are primarily cosmic ray hits or bad pixels identified by the pipeline. Next, we collapsed the image along the wavelength axis and calculated the centroid position of the spectral trace by fitting a Gaussian to the flux profile and retrieving the peak location, rounded to the nearest integer pixel. We then established two regions centered on the centroid row: (1) the spectral extraction region, extending L1 pixels from the centroid, within which the irradiance is measured (i.e., a rectangular box of height 2L1+1), and (2) the background region, which begins at L2 pixels from the centroid row and extends to the edge of the subarray. For each column k, we applied a window spanning (k-W,k+W) and constructed a template PSF by subtracting the median flux level within the background region and computing the median column flux array within the window. We then fit this template PSF to the flux values in column k using a standard least-squares algorithm, with a multiplicative scaling factor and an additive background level. The fitting procedure was carried out iteratively, each time masking $5\sigma$ outliers. The resultant extracted flux for column k is the integrated value of the best-fit scaled template PSF. After the spectrum was computed at each of the two dither positions, we applied a 20-pixel-wide moving median filter to trim $3\sigma$ outliers and averaged the two spectra together to arrive at the final spectrum.

This PSF fitting method naturally accounts for the small subpixel shifts in centroid position across the spectral trace and the changing cross-dispersion spectral profile shape, while providing increased precision over traditional rectangular aperture extraction. We experimented with various settings for L1, L2, and W and found that the resultant spectrum was largely unchanged across a broad range of values. For the spectrum presented in this paper, we used L1 = 5 px, L2 = 5 px, W = 20 px. To ensure that our rectangular extraction region was wide enough to fully enclose the spectral trace across the entire wavelength range, thereby preventing wavelength-dependent sampling biases, we ran a series of spectral extractions with different values of L1 and found that the resultant integrated fluxes plateau beyond L1 = 3 px.

4.2. MIRI Data Reduction

For the MIRI observations, our data reduction procedure was largely analogous to the NIRSpec spectral extractions. After passing the uncalibrated data from both the science and background observations through Stage 1 of the JWST pipeline (converting raw pixel counts to count rates), we collated the science and background rate files into associations, sorted by sub-band and channel, before proceeding with Stage 2. We manually turned on the background step, which carries out pixel-wise subtraction on each pair of science and background images within an association in order to remove the thermal background and zodiacal light, as well as mitigate hot pixels, column striping from variable dark current, and residual flat field artifacts. MIRI MRS data suffer from significant fringing, caused by internal reflection in the detector. As part of the Stage 2 processing, two subroutines are available to correct the fringes: *fringe* and *residual_fringe*, the first of which runs by default, while the latter is manually activated in our data processing code.

The output of Stage 2 is flux-calibrated 3D IFU data cubes consisting of a stack of 2D spatially-rectified images (i.e., image slices), each corresponding to a different wavelength. The x and y pixel coordinates of the centroid were calculated by collapsing the cube along the wavelength axis and fitting a 2D Gaussian to the target. There are several challenges associated with spectral extraction of the MIRI MRS data cubes. First, while the Stage 2 processing significantly reduces the fringing across the detector, the effects are not entirely removed, resulting in low-level flux variations in the image slices. Consequently, there are oscillations in the shape of the source PSF as a function of wavelength that can lead to biases when creating median-averaged template PSFs in an analogous manner to our NIRSpec spectral extraction. Indeed, when comparing the spectra derived using simple aperture extraction and PSF fitting, we observed systematic deviations in the wavelength regions where the residual fringing is most apparent – typically at the short- and long-wavelength ends of each sub-band. The PSF fitting extraction produced oscillatory flux modulations that are not present in the simple aperture extraction, yielding higher levels of correlated noise in the spectrum on wavelength scales comparable to the characteristic fringing. The relative deviation between the two different extraction methods was also found to vary with the choice of moving window width (W) used for constructing the template PSFs. Therefore, we chose simple aperture extraction for producing the MIRI spectra presented in this paper.

The second challenge for MIRI MRS spectral extraction is the large spatial extent of point-source PSFs. As can be seen in Figure 2, the PSF consists of a central peak and six secondary diffraction lobes at wider separations that contain a nonnegligible fraction of the total dispersed flux. The dimension of the image slices and the corresponding fraction of the source

PSF that falls outside of the field of view vary with channel, dither position, and wavelength. We selected a relatively small 7x7 pixel square aperture centered on the centroid pixel for our spectral extraction, which ensured that every pixel within the aperture is illuminated at all wavelengths, and carried out flux corrections at a later stage (described below). Larger apertures caused some of the edge pixels to fall outside the illuminated portion of the field of view in some image slices, while smaller aperture sizes yielded increased correlated noise in the extracted spectra, which we attribute to the effects of the undersampled PSF at the pixel scale of the MIRI MRS IFU. We also experimented with circular apertures with wavelength-dependent sizes that vary with the changing size of the source PSF, but also found slightly increased correlated noise, which are likely due to subpixel interpolation artifacts from the undersampled PSF.

While the Stage 2 processing removed most of the background, there remained a small nonzero median flux level at wide separations from Didymos. To remove the remaining background we subtracted the median value of pixels lying more than 15 pixels from the centroid prior to summing the flux within the aperture. Varying the size of the background region by as much as 5 pixels in either direction yielded shifts to the extracted flux that were smaller than the flux uncertainties; likewise, skipping the background subtraction altogether did not affect the resultant spectrum, since typical values of the residual background pixel flux level are <0.1% of the PSF peak value. The spectral extraction and background regions are indicated in Figure 2.

Even after applying the Stage 2 defringing steps in the JWST pipeline, clearly discernible fringes remain in the extracted spectra. To mitigate these, we ran the spectra through the *fit_residual_fringe_1d* subroutine in the JWST pipeline, which takes as input the 1D extracted spectra, masks out significant spectral features, removes the continuum trend, and runs a sinusoidal fit to residual array that is subsequently subtracted from the initial spectrum. This step significantly improved the quality of the spectra, particularly in Channels 1 and 2, where the amplitude of the fringes was reduced by a factor of 2-4. After carrying out outlier trimming in a manner identical to the NIRSpec spectra, we calculated the mean of the spectra from the four dithers to arrive at the combined MIRI MRS spectrum.

The wide extent of the MIRI PSFs, specifically the secondary diffraction pattern, leads to significant flux losses outside of our 7x7 pixel extraction aperture. In our analysis, we corrected for these biases by benchmarking the relative flux loss to observations of a standard star. We selected the A6V star HD 163466 (V = 6.85 mag), which was observed as part of the Cycle 1 flux calibration program 1536 (PI: Karl Gordon). The uncalibrated data from the standard star observation were run through the same processing procedure as our Didymos observations. We then extracted the star's spectrum using the same 7x7 pixel aperture and compared it to the available CALSPEC spectrum (Bohlin et al. 2014). By dividing the two spectra in each MIRI channel and sub-band and fitting second- or third-order polynomials to the resultant ratio curves, we empirically quantified both the relative flux loss outside of the extraction aperture and any systematic issues with the current flux calibration in the JWST pipeline. Figure 3 shows the ratio curves that we derived. The ratio decreases with increasing wavelength due to the increasing size of the source PSF and the corresponding increase in the fraction of flux that falls outside of the fixed 7x7 pixel aperture. The discontinuity between sub-bands 2C and 3A reflects the different pixel scales of the detectors (0.13" for Channels 1 and 2, 0.20" for Channels 3 and 4). We divided these ratio curves from the extracted Didymos spectrum to recover the full irradiance of the target.

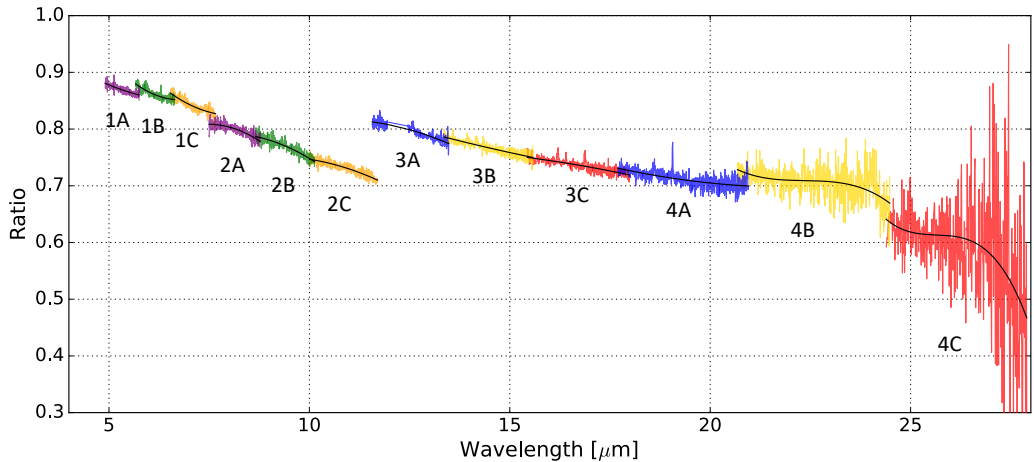

Figure 3: Ratio of the extracted MIRI MRS spectrum of the A-type standard star HD 163466 from a fixed 7x7 pixel square aperture to the CALSPEC spectrum. The individual sub-bands are labeled, and the black curves show low-order polynomial fits to the ratio curves, which were used as a benchmark for correcting Didymos' extracted spectrum for flux losses outside of the aperture.

We note that the calibrations for MIRI continue to evolve. Recent updates to the pipeline[1] incorporated improved flux calibration from in-flight data and corrections for the observed time-dependent throughput loss. However, due to the low flux of standard stars at the longest wavelengths, particularly Channel 4 (17.7—27.9 µm), the absolute flux levels in those regions are not considered reliable (David Law, private communication, also Figure 4). Indeed, we found that the continuum shape of the spectrum beyond 20 µm shows unphysical curvature that is not well-matched by any thermal model. An artifact near 12.2 µm manifests as a bump in flux and is due to an imperfect correction of a second-order flux leak from the 6 µm region that was incorporated into the initial flux calibration response curve in earlier versions of the JWST pipeline. Given the evolving calibration for Channel 4 and the large discrepancy between the measurements and model at long wavelengths, we do not further report or interpret data at wavelengths longer than 20 µm in this work, and we ignore or call out the regions with known artifacts as appropriate.

There are significant systematic flux level differences between the grating settings that can be attributed to the rotational variation of Didymos between the observations, which spanned a significant fraction of Didymos' rotation (Table 1). In our analysis, we assume that the composition, and therefore emissivity spectrum, of Didymos is uniform across the surface, with the variations in the flux across its orbit due solely to the changing sky-projected area of the target between the observations. Having ensured that our individual sub-band spectra are properly scaled to capture the full source flux during each exposure, we simply renormalize adjacent sub-bands to match across the overlapping regions. We corrected for the flux offsets across the full MIRI MRS spectrum by normalizing each sub-band's spectrum relative to the flux level in sub-band 2C (10.0—11.7 μm); the required normalization factors are at most 5%. The normalization factors for sub-bands obtained simultaneously with the same grating setting (e.g., 1B, 2B, 3B) are consistent to well within 0.5%, which serves as empirical validation for our assumption of a uniform emissivity spectrum across Didymos' surface.

---

[1] https://www.stsci.edu/contents/news/jwst/2023/miri-mrs-data-processing-improvements-are-now-available

## 5. Thermal properties

To study the composition of Didymos, its thermal flux must be removed from the NIRSpec and MIRI spectra. For the NIRSpec data we use a thermal modeling code that implements a version of the Standard Thermal Model (STM; Lebofsky et al. 1986) modified to have some features of the Near Earth Asteroid Thermal Model (NEATM: Harris 1998). All of the required inputs to the STM are known, other than the "beaming parameter" (η), which we treat as a free parameter rather than fixing it at the STM default value of 0.756. We use the albedo (0.15 ± 0.02) and size information for Didymos from Daly et al. (2023). The value of η used for the NIRSpec data is 1.93, consistent with what is found for similar-sized objects at similar phase angles (Delbo et al. 2007). The more recent NEATM is commonly used for small and/or near-Earth asteroids. The NEATM differs from the STM by allowing η to float and by using a wavelength- and phase-angle-dependent value for the infrared phase coefficient to account for non-zero nighttime thermal emission. The code we use already treats η as a free parameter, as NEATM does. The difference between using the default wavelength-independent STM infrared phase coefficient of 0.01 magnitude per degree and a wavelength-dependent value for the normalized NIRSpec data is estimated to amount to < 0.5% across the entire wavelength range of data presented here, a level of uncertainty that is much smaller than the reported uncertainty in η and data scatter.

For the MIRI data we use the Advanced Thermophysical Model (ATPM; Wolters et al. 2011, Rozitis & Green 2011, Rozitis et al. 2020) along with the latest shape models and orbit information for Didymos and Dimorphos (Daly et al. 2023, Thomas et al. 2023, Barnouin et al. in prep). This model calculates the temperatures and thermal fluxes for each facet of an object's shape model, outputting the predicted spectral energy distribution (either globally or locally) for wavelengths of interest. There is no η used in thermophysical models because the factors abstractly represented by η like thermal inertia, shape, and roughness are explicit inputs or values that are fit by these models. Rozitis et al. (2020) contains a fuller description of the ATM.

Figure 4 shows the flux from Didymos in Jy as measured by MIRI, along with the best ATM fit. Given the issues with Channel 4 discussed above, especially at the longer wavelengths, we do not consider or show the problematic wavelengths > 20 μm in this work, and have shaded the 17.7-20 μm region in figures to denote Channel 4 data.

While NIRSpec and MIRI have overlapping wavelength coverage near 5 µm, we do not construct a single reflectance spectrum over the 0.6–20-µm range. Such an undertaking would require appropriately scaling the NIRSpec and MIRI data to account for their different distances from the Sun and JWST, different phase angles, and lightcurve effects, and doing a simultaneous fit to those data including different scattering and emission functions. Again, given the evolving JWST instrument calibrations, such an effort would be premature. We therefore separately discuss the NIRSpec and MIRI results in Sections 6.1 and 6.2, respectively, before synthesizing those discussions in Section 6.3.

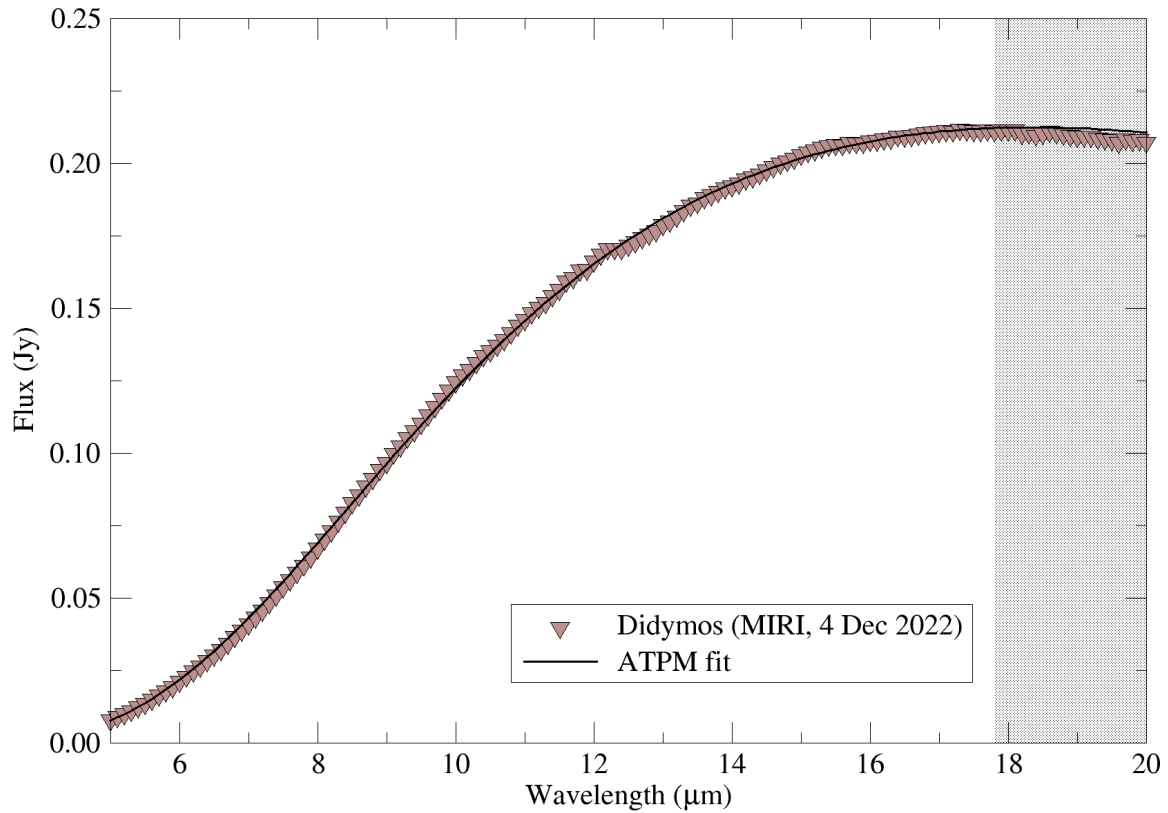

Figure 4: Spectral Energy Distribution for Didymos from MIRI compared to the best-fit ATM. The shaded region covers Channel 4 of MIRI, which currently has some unresolved calibration problems.

We can use the ATM fits to provide estimated thermal inertias from the MIRI and NIRSpec data of 260 ± 30 J m$^{-2}$ K$^{-1}$ s$^{-1/2}$ and 290 ± 50 J m$^{-2}$ K$^{-1}$ s$^{-1/2}$, respectively. These thermal inertia measurements are fully consistent with one another and consistent with what is seen on other objects of Didymos' size (760 m diameter): the thermal inertia of Bennu (500 m) is 300 ± 30 J m$^{-2}$ K$^{-1}$ s$^{-1/2}$ (Rozitis et al. 2020), while that of the smaller Itokawa (330 m) is 700 ± 200 J m$^{-2}$ K$^{-1}$ s$^{-1/2}$ (Müller et al. 2014). Figure 5 compares the NIRSpec-derived thermal inertia of Didymos to that of other asteroids < 5 km in diameter, including 6 other S-complex objects and 14 asteroids of other or unknown spectral classes, with data taken from the MacLennan and Emery (2021) compilation.

Gundlach and Blum (2013) developed a method to estimate particle sizes in airless body regoliths given thermal inertia, sub-solar temperature, and typical thermal properties for asteroidal minerals. Using the estimated thermal inertias discussed above and a range of estimated sub-solar temperatures at the time of the JWST measurements (300-350 K), and considering a range of packing fractions from 0.1—0.6 (in other words, a regolith porosity of 0.4—0.9) results in an estimated particle size of 2-7 mm on Didymos. For comparison, using the same method Gundlach and Blum (2013) report particle sizes of roughly 10-25 mm on Itokawa, 100 µm-2 mm on 1998 WT24, 4-40 mm on 1999 JU3, and 300 µm-2 mm for 1996 FG3. These four objects were the only sub-km objects with reported particle sizes, and Didymos' calculated particle size falls within the range they define.

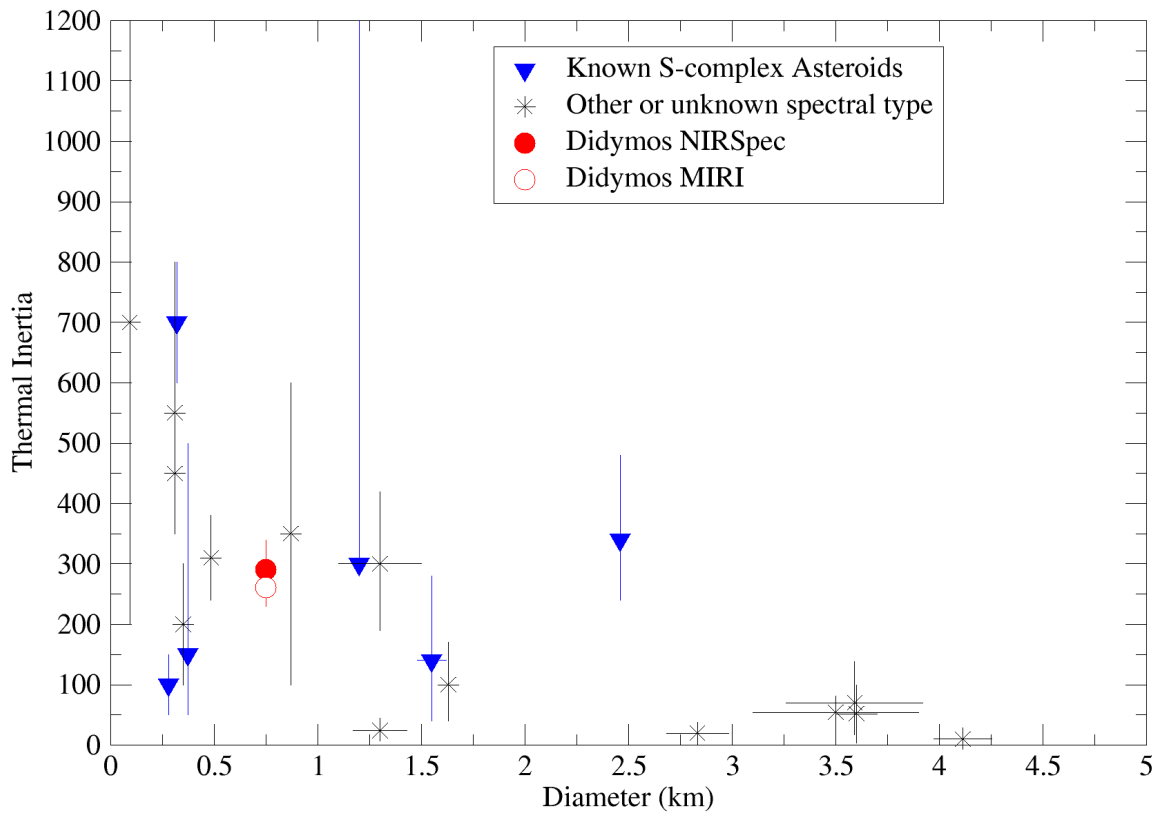

Figure 5: The thermal inertia of Didymos (open and closed red circles) is similar to that of other measured thermal inertias for asteroids < 5 km in diameter. There does not appear to be a compositional dependence on thermal inertia that can be discerned from this sample. The data for asteroids other than Didymos were compiled from the literature by McLennan and Emery (2021), with original citations found in that work.

## 6. Composition

6.1 Visible-Near-Infrared

Figure 6 shows the NIRSpec 0.75—2.5-µm reflectance spectrum of Didymos compared to a spectrum of Didymos obtained by XSHOOTER on the VLT on 27 September 2022 (de León et al in prep). The phase angle for the XSHOOTER data was 53°, a coincidental match given Didymos' phase angle increased to > 70° from both the Earth and JWST after the XSHOOTER observations before decreasing to 54° for the NIRSpec observations. Absorption bands near 1 and 2 µm are evident, and the spectrum is consistent with previous observations in this wavelength range. The formal error bars for the NIRSpec observations are smaller than the datapoints, but the true uncertainty is better represented by the ~2-3% scatter from point to point. We normalize these spectra to 1.5 µm near a reflectance peak rather than a more typical 700 or 750 nm in order to avoid involving possibly discrepant points near the edge of the NIRSpec detector. The NIRSpec spectrum appears to have a deeper 1-µm absorption band than the XSHOOTER spectrum, but whether this is a true difference in the spectra or a result of discrepant points and scatter near the edge of the NIRSpec detector is unclear. When normalized to 1.5 µm, the general agreement between the two datasets from 0.83—2.5 µm is excellent.

Figure 6 also includes the average S-class and Q-class asteroid from DeMeo et al. (2009), also normalized to equal 1 at 1.5 µm. The match between the average S and the Didymos spectra is evident, confirming earlier classifications (de León et al. 2006). The mismatch between Didymos and the average Q-class spectrum is consistent with the general paradigm that S-class asteroids are space-weathered versions of Q-class objects (Binzel et al. 2004, Hiroi et al. 2006).

Figure 7 compares Didymos to the Chelyabinsk meteorite in the 0.75—2.5-µm region. We choose Chelyabinsk as a recent LL chondrite fall and to allow comparison of Didymos to both typical and shock-darkened ordinary chondrite spectra– Didymos has been seen to have not only a lower-than-typical albedo for S asteroids but evidence of albedo units, and shocked material may be present within boulders on Dimorphos (Sunshine et al. 2022).

Four Chelyabinsk spectra from the RELAB database (Milliken 2020, Milliken et al. 2021, Jenniskens PI of data) are shown in Figure 7, representing the light-colored and dark lithologies, both as cm-sized chips and as powders with particle sizes < 125 µm. The NIRSpec spectrum is most closely visually matched by the light-colored Chelyabinsk powder, though the light-colored chip differs in spectral slope rather than absorption band depth or position. The dark-colored Chelyabinsk samples are poor matches due to their lack of 1- and 2-µm absorption bands.

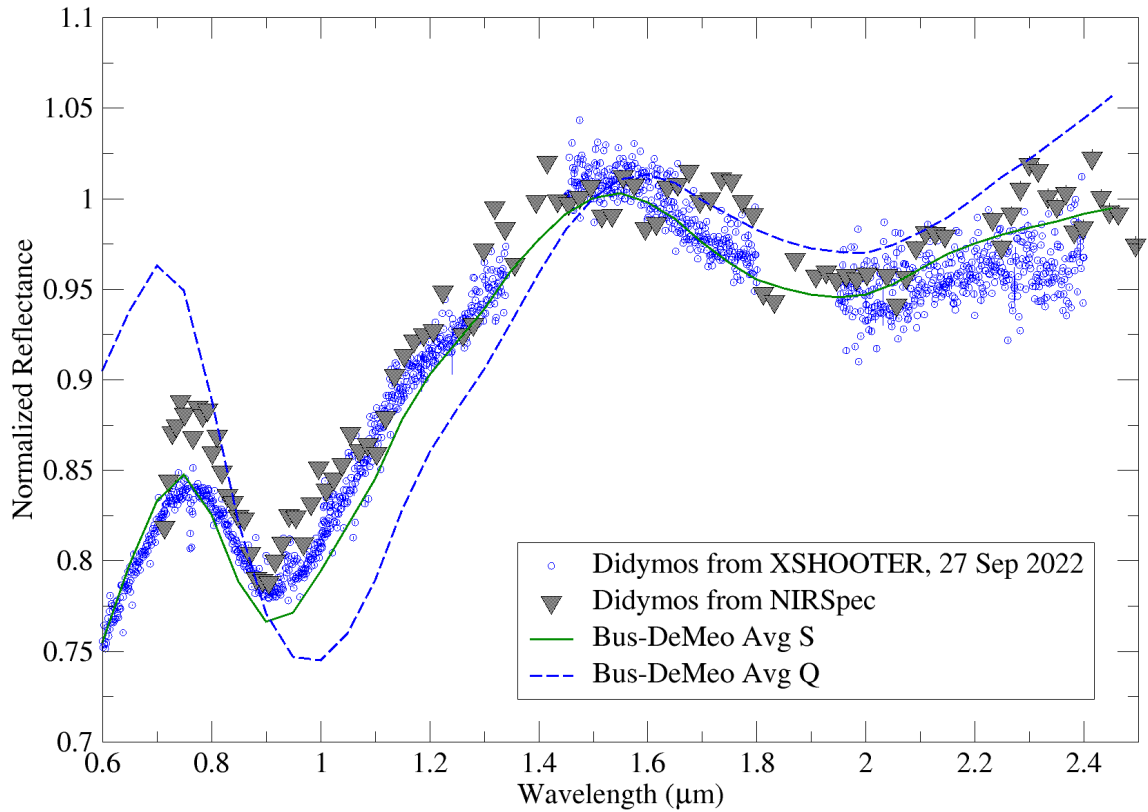

Figure 6: Comparison of Didymos NIRSpec data and XSHOOTER data from 27 September 2022 (de Leon et al. in prep), scaled to match at 1.5 μm. The overall agreement is excellent. The uncertainties in the NIRSpec data are not well-represented by the formal error bars, which are smaller than the datapoints, but are better represented by the point-to-point scatter. Also shown are the average S-class and Q-class asteroids from DeMeo et al. (2009). Didymos is best matched by the S-class spectrum.

Silicate compositions can be estimated for S-complex asteroids from their near-infrared spectra, and methods for doing so have been evolving for over 30 years (Gaffey et al. 1993, Dunn et al. 2010, McClure and Lindsay 2022a, McClure and Lindsay 2022b). The most common compositional determination scheme involves measurements of the center of "Band 1" (an absorption band near 1 μm associated with olivine, pyroxene, or both) and the ratio of the area of "Band 2" (an absorption band near 2 μm associated with pyroxene) to the area of Band 1. We conservatively estimate the Band 1 center position in the NIRSpec spectrum as 0.94 ± 0.01 and the Band Area Ratio (BAR) as 0.4 ± 0.1 following the practice of de Leon et al. (2010). For comparison, the Band 1 center and BAR values for the Chelyabinsk light powder spectrum in Figure 5 are 0.940 ± 0.005 μm and 0.76 ± 0.02, respectively. These band parameters place both Didymos and Chelyabinsk squarely in the Ordinary Chondrite (OC) region in the original Gaffey et al. (1993) framework as well as the McClure and Lindsay (2022a) framework.

Dunn et al. (2010, 2013) provided means of estimating the olivine and pyroxene compositions and the ol/(ol+px) ratios of OC meteorites and their asteroid analogs. The BAR value for Didymos, even with these relatively large uncertainties, is most consistent with LL-type OCs (Dunn et al. 2010) and suggests an ol/(ol+px) ratio of 0.57 ± 0.16. The Chelyabinsk BAR, however, is more typical of H or L chondrites than LL chondrites. The Band 1 centers for both Chelyabinsk and Didymos are more consistent with H or L chondrites than LL chondrites. Given the ambiguity of these results, and given the visual similarity between Didymos and Chelyabinsk (known to be an LL chondrite from laboratory studies) in Figure 7, we suggest that the NIRSpec spectrum of Didymos is overall most consistent with L/LL chondrites in the 0.6—2.5-μm region, in agreement with the findings of Dunn et al. (2013) and Ieva et al. (2022).

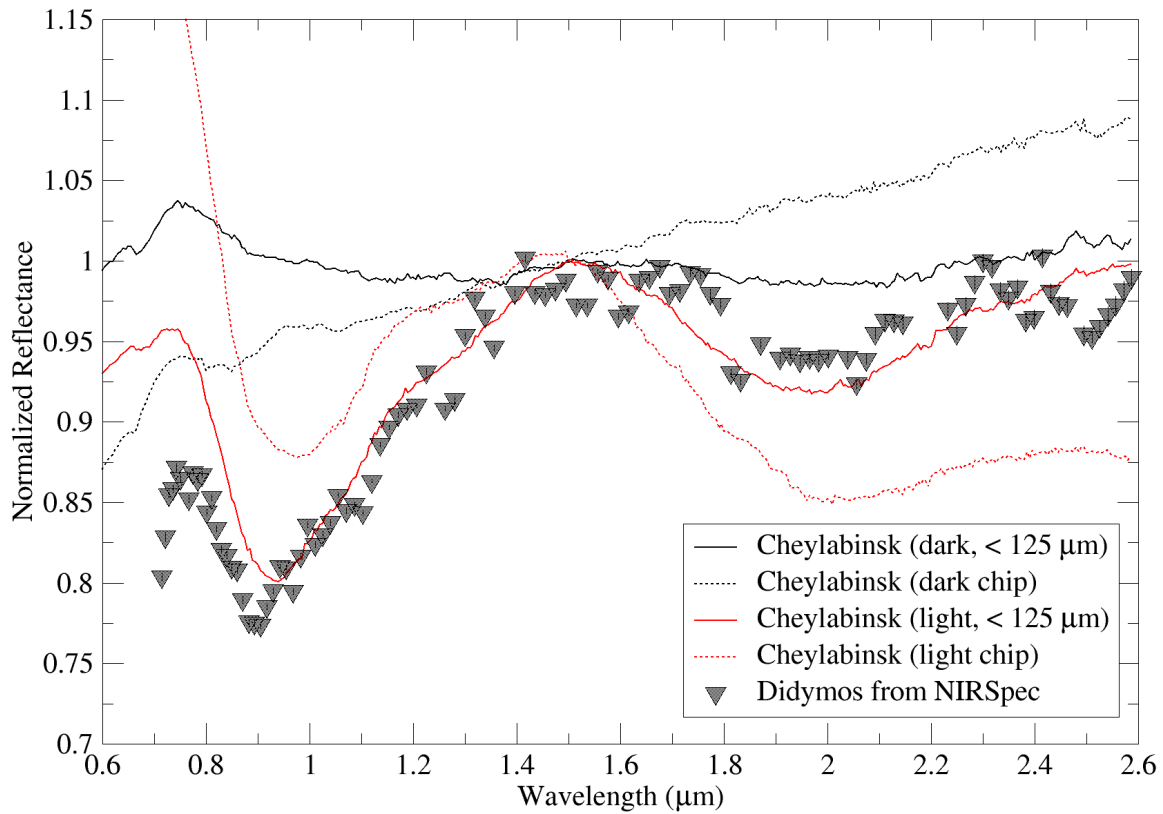

Figure 7: Comparison of the NIRSpec data to RELAB spectra of Chelyabinsk (Jenniskens PI), normalized to 1.5 μm. Didymos is a closer match to the powdered light-colored lithology of Chelyabinsk than a chip of the light-colored lithology or either of the dark lithology spectra, suggesting cm-sized particles and shock-darkened regions are not important contributors to the overall reflectance spectrum at hemispheric scales in the Didymos system.

Absorption features near 3 μm are diagnostic for hydrated and hydroxylated minerals on airless surfaces at near-Earth asteroidal temperatures. Hydrated carbonaceous chondrite meteorites typically have band minima near 2.7—2.8 μm due to phyllosilicates (Beck et al. 2010, Takir et al. 2013, 2019, Bates et al., 2021), and similar band shapes are commonly seen in low-albedo asteroids (Rivkin et al., 2015, Hamilton et al. 2019). Band centers at longer wavelengths (~3.1 μm) have also commonly been seen on main-belt asteroids, and attributed to ice and/or ammoniated minerals (King et al. 1992, Takir et al. 2012, Rivkin et al. 2022). Other minerals, including (but not limited to) goethite (Beck et al. 2011) and the clay mineral cronstedite (Rivkin et al. 2006), can also have band minima at wavelengths near 3.0-3.1 μm. Unlike the 0.6—2.5-µm wavelength portion of the Didymos data, the 2.5—5-µm spectroscopic data were unobtainable from any facility except JWST.

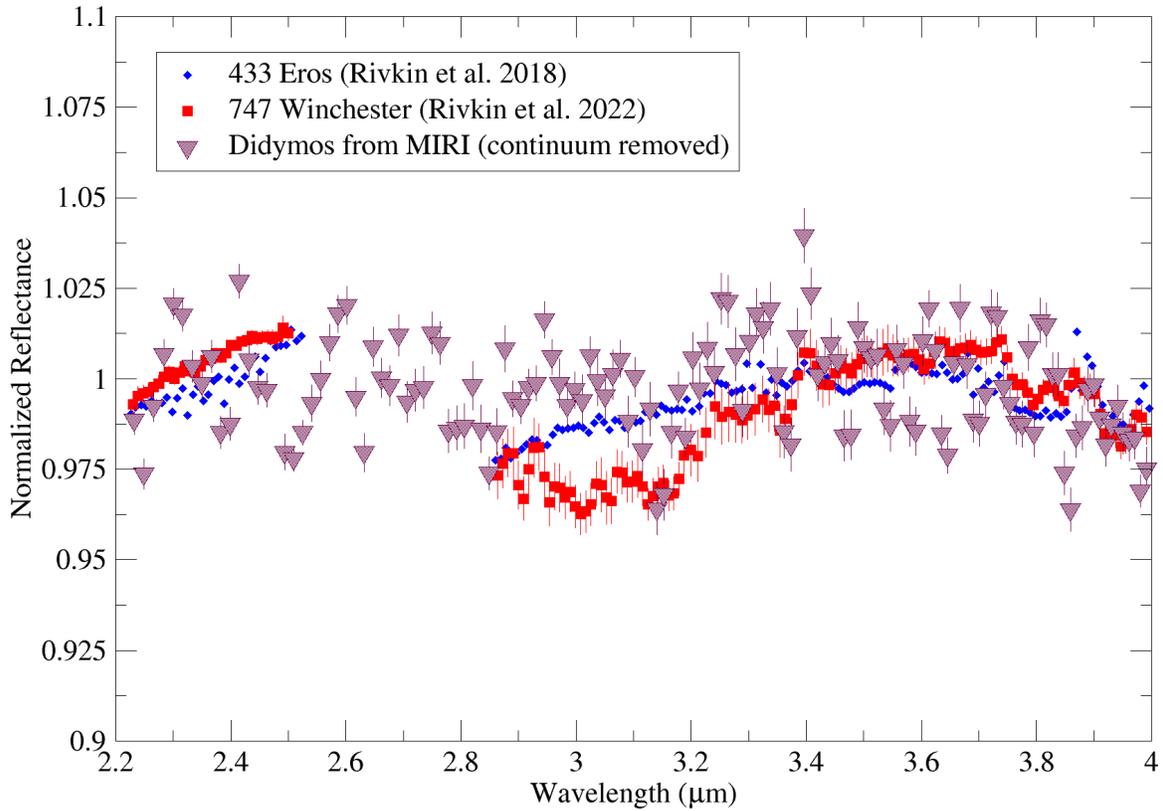

Figure 8: Didymos compared to Eros from Rivkin et al. (2018) and the low-albedo asteroid Winchester from Rivkin et al (2022). Eros has a discernable absorption in the 3-µm region, interpreted as due to hydrated minerals with an estimated concentration of a few hundred ppm of hydrogen. The uncertainties on the Eros spectrum are smaller than the data points when they are not seen. Didymos shows no indication of an absorption in this region, consistent with an anhydrous surface. Because the scatter is large enough to encompass the Eros spectrum, we suggest that a possible upper limit on hydration on Didymos is similarly a few hundred ppm of water.

The 2.2—4.0-µm NIRSpec spectrum of Didymos is shown in Figure 8, along with a spectrum of (433) Eros (Rivkin et al. 2018), with both normalized to 2.4 µm. This spectrum of Eros has been interpreted as having a shallow absorption with a band depth ~1-2% near 2.9 µm. Didymos shows no discernable absorption band, though the scatter in its spectrum is large enough to hide an absorption of the depth seen on Eros, in principle. Unlike the ground-based Eros spectrum, however, the NIRSpec data covers the 2.5—2.8-µm region where absorption due to $OH^-$ is strongest. Here too, there is no discernable absorption band in the Didymos spectrum, and the pattern of lower and higher data points are not what is typically seen in hydrated mineral absorption bands. Also included is a normalized spectrum of the outer main-belt C-class asteroid (747) Winchester, from Rivkin et al. (2022). Winchester shows an example of the second, common shape for 3-µm absorption bands on asteroids, centered at 3.0—3.1 µm. As with Eros, the absorption band on Winchester is several tenths of μm in width, in contrast to the higher-frequency scatter seen in Didymos' spectrum.

Rivkin et al. (2018) estimated that Eros had a water concentration of a few hundred ppm based on its 3-μm band depth and comparisons to data from Vesta. Given the scatter in Didymos' spectrum, we estimate a similar value as an upper limit on water concentration in the minerals on Didymos' surface but again note that there is no discernable absorption band and the *lower* limit of water concentration is consistent with zero.

6.2 Mid-Infrared
Salisbury et al. (1991) discusses analysis of the mid-infrared spectra of asteroidal materials based on Restsrahlen bands, absorption bands, Christiansen features (CF), and transparency features (TF). Interpretation of the emissivity spectrum of Didymos in Figure 9 shows a CF apparent near 8.6--8.8 μm, though the uncertainties in the MIRI data make the exact location of the CF difficult to precisely measure. Reststrahlen bands can be seen at roughly 9—12 μm and 14.5—17.5 μm.

The 10 um Si-O stretch fundamental mode in silicates is sensitive to grain size (Hunt & Logan 1972). Bramble et al (2021b) reported Transparency features (TF) centered near 12.5 μm and stretching from 11.8–13.3 μm in measurements of ordinary chondrites and OC-like mixtures in asteroid-like conditions (Figures 10, 11). This wavelength region in the MIRI spectrum unfortunately includes the artifact at 12.2 μm mentioned in Section 4.2, but the emissivity spectrum is consistent with a TF that is deepest near 12.8 μm. A second emissivity peak is present near 15.4 μm (Figure 9). The interpretation of the emissivity at the shortest MIRI wavelengths is still uncertain (see Section 6.2.2) but shows an apparent TF for Didymos stretching from roughly 6.0—8.2 μm.

The emission from planetary surfaces, and thus the position and strength of features just mentioned, is affected by factors including grain size, near-surface temperature gradients, atmospheric pressure (and its absence) and porosity (Urquhart and Jakosky 1997, Reddy et al. 2015). There has not yet been a comprehensive study in midinfrared wavelengths of the emissivity spectra of solid solutions like olivine and pyroxene in vacuum conditions, complicating matters. Measurements by Shirley and Glotch (2019) and Bramble et al. (2021a) show that the CF of minerals shifts ~0.05—0.2 μm shortward in vacuum conditions compared to ambient conditions, with the exact amount again dependent upon grain size, temperature, and composition. Quantitative matching between mid-infrared spectra of asteroids and laboratory spectra, particularly when particle sizes are close to the wavelength of light, remains an active research topic, and one whose overall solution is beyond the scope of this work.

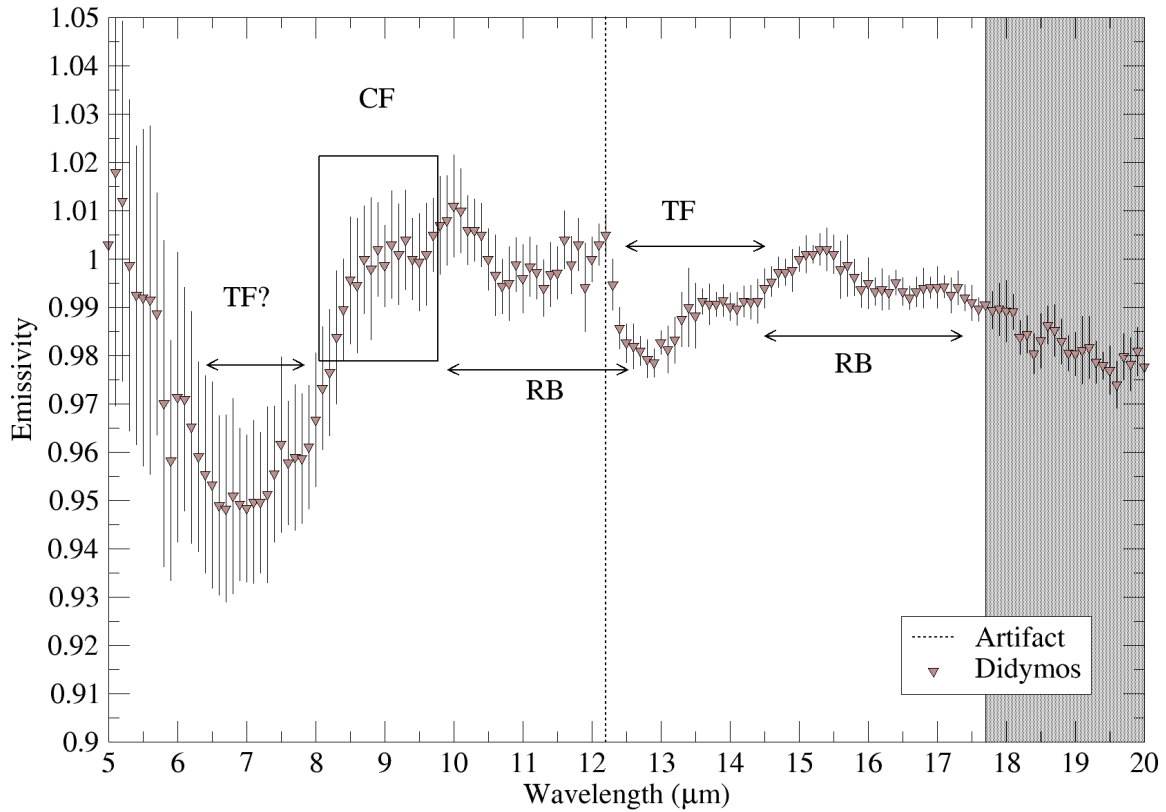

Figure 9: MIRI emissivity spectrum as calculated using the ATM (section 5). The shaded area is affected by limitations in the JWST calibration, as discussed in the text, and the artifact position near 12.2 µm is shown. Also shown are the interpreted positions of the Christensen feature (CF), Reststrahlen bands (RF) and possible location of Transparency features (TF). The precise CF location is difficult to determine given the observational uncertainties.

### 6.2.1 Mid-IR Comparison to Ordinary Chondrites

We can take advantage of the association of Didymos with ordinary chondrite meteorites from the NIRSpec data and previous work (de León et al., 2006; Dunn et al., 2013), and turn to laboratory measurements of these materials made in vacuum for insight. Bramble et al. (2021a, 2021b) measured the thermal emission of forsterite, enstatite, metal, and plagioclase minerals, along with mixtures representing the bulk composition of H, L, and LL chondrites, in simulated asteroidal conditions and at a variety of particle sizes.  Bramble (2020) provided similar measurements for 5 ordinary chondrites. Figure 10 compares Didymos' MIRI spectrum to the L-chondrite-like mixture measured for several different particle size ranges, with the laboratory spectra normalized to equal the Didymos spectrum near 15 µm.  While these mixtures do not match OC compositions in detail (for instance, the olivine in OC meteorites is not pure forsterite and the low-Ca pyroxene is not pure enstatite), the comparison is instructive for constraining Didymos' overall consistency with OC meteorites. The drop in Didymos' emissivity shortward of 8 µm and the lack of a similarly steep drop longward of 15 µm are better matched by the finest (<25 µm) size fraction of the L-chondrite-like mixture vs. coarser-grained mixtures. The presence of a TF near 13 µm in the MIRI data is consistent with the finest fraction, which also has this feature, as opposed to the coarser fractions, which lack the TF. A dashed vertical line shows the location of the artifact discussed in Section 4.2.

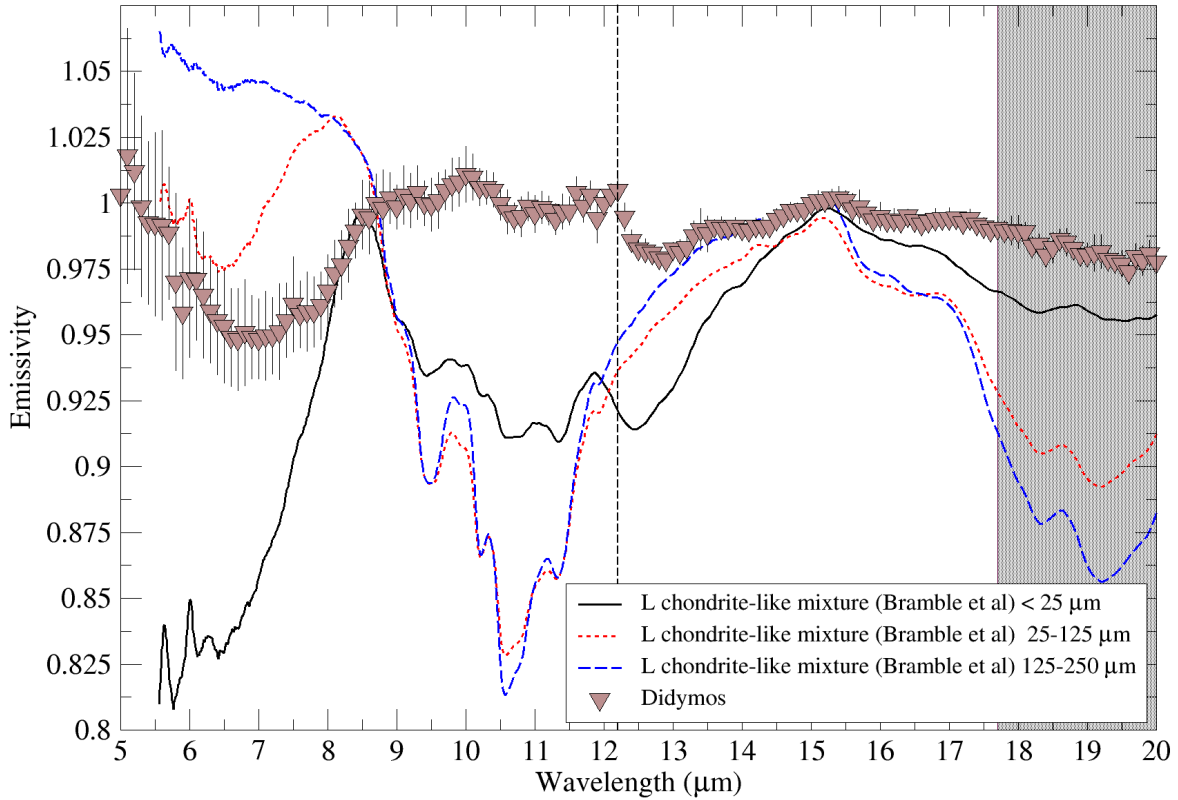

Figure 10: Comparison of emissivity spectra, normalized at their 15-µm emissivity peaks, of L-chondrite analog mixtures (from Bramble et al. 2021b) to Didymos. The finest fraction of the analog mixture is qualitatively similar to the Didymos spectrum across much of the wavelength range, though the asteroid emissivity features are generally muted longward of ~8 µm. The lack of an emissivity drop longward of 8 µm in Didymos' spectrum could be due to differences in porosity, hyperfine particles, or other physical causes vs. compositional ones. The coarsest fraction of the analog is a poor match across the wavelength range, and only the finest fraction shows a TF like Didymos does.

The mixtures all show an emissivity drop from 9—14 µm that is not seen in Didymos' spectrum. Work by Vernazza et al. (2012) suggested that mid-infrared emission peaks in asteroid spectra like what is seen in Didymos' spectrum are likely due to very high porosity (>90%) regolith. Recent work by Sultana et al. (2023) suggests that high porosity alone cannot be the cause, instead suggesting that hyperfine particles mixed with opaque materials can cause such peaks. Space weathering is seen to reduce spectral contrast in the Reststrahlen band and transparency feature region (Shirley et al. 2023), which is consistent with what is seen in the Didymos MIRI spectrum and consistent with Didymos being a space-weathered object.

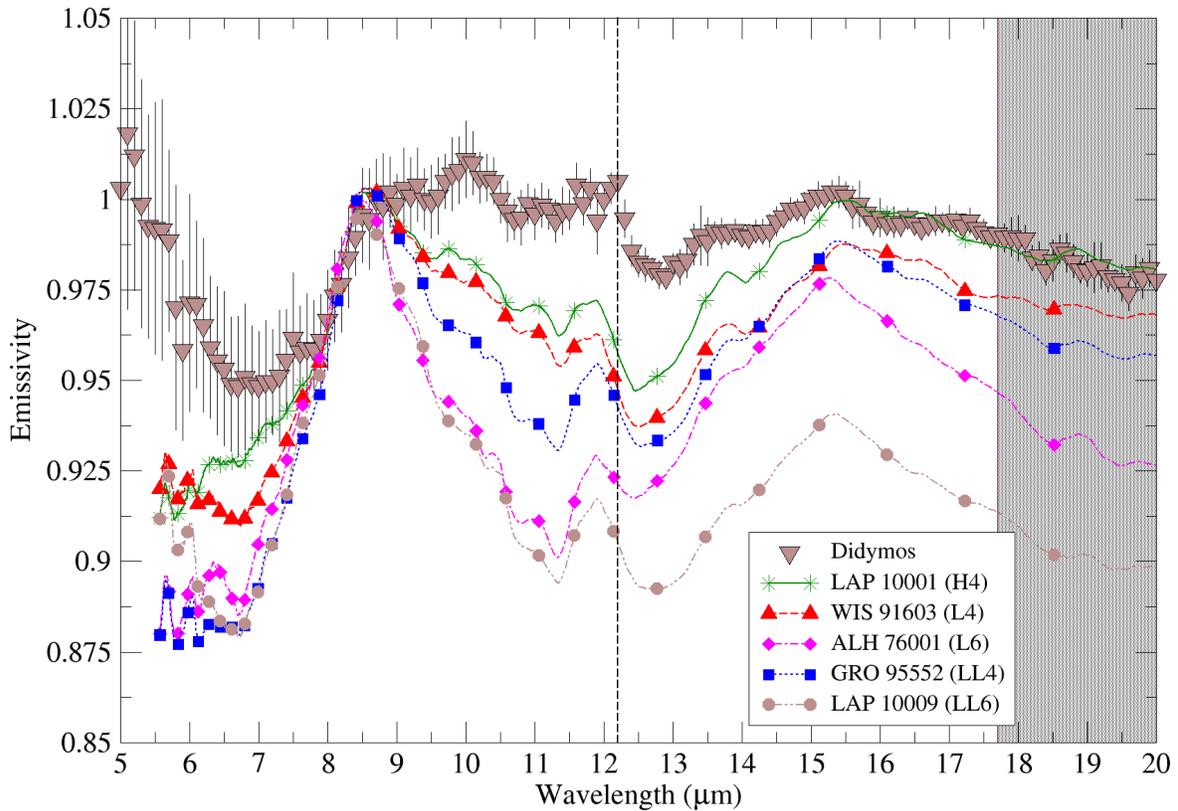

Figure 11 MIRI spectrum of Didymos compared to the finest fraction of several OC meteorites from Bramble (2020) measured in simulated asteroidal conditions. The spectra are all qualitatively similar, though the Didymos spectrum has emissivity features near 10 µm and 17 µm not present in the OC spectra, and has generally higher emissivity than them at λ > 9 µm. The position of the ~15-µm peak is more consistent with the LL chondrites than the other chondrites, though the number of available comparison spectra is small.

Figure 11 compares spectra of the Didymos MIRI spectrum to the finest size fraction of H, L, and LL chondrites, measured in a simulated asteroidal environment (SAE) by Bramble (2020). These spectra are all normalized to match the Didymos spectrum near 7.5—8.5 µm. The OC spectra are qualitative matches to Didymos' spectrum in ways similar to the finest fraction of the L-chondrite simulant shown in Figure 10. The OC spectra mismatch Didymos in the 9—14-µm region in the same way as the simulant spectrum does, although the mismatch is generally smaller for the OC spectra. Additionally, a prominent peaks is seen near 10.5 µm in the Didymos spectrum that is not seen in the OC spectra. These are discussed in Section 6.2.2. The average CF position for the fine-grained SAE OC samples, averaged over measurements taken at various temperatures, from Bramble (2020) is 8.56 ± 0.06 µm, somewhat shortward of what appears to be the CF position for Didymos by inspection but within observational uncertainty. Glotch et al. (2015) concluded based on studies of lunar swirls that space weathering of lunar materials causes a shift of 0.08-0.09 µm in CF position toward longer wavelengths. The observational uncertainty in the MIRI dataset does not allow us to robustly assess whether a shift of that magnitude is present compared to the OC spectra, although it does appear potentially consistent. The position of the ~15-µm peak in the Bramble (2021) SAE OC spectra, again averaged over various temperatures, is 15.44 ± 0.11 µm, a good match to the peak seen in the Didymos spectrum. The position of the 15-µm emissivity feature in Didymos is best matched by the LL chondrite spectra, with vs. the other meteorite groups, but the sample size is limited.

Salisbury et al. (1991) measured the mid-infrared spectra of 60 meteorite samples, including 19 OC meteorites. The CF position for Didymos (8.6—8.8 µm) is most consistent with the range seen in LL chondrites by Salisbury et al. (8.68—8.81 µm), although they are also potentially consistent with the range seen in H chondrites (8.60—8.78 µm) and L chondrites (8.58—8.74 µm) in that work. We also note that the Salisbury et al. (1991) measurements were made in ambient conditions rather than vacuum and a direct comparison to the CF position of Didymos, measured in vacuum but potentially space weathered, may not be completely appropriate.

### 6.2.2  Comparison to minerals

The mineralogy of ordinary chondrites is dominated by olivine and pyroxene, with these minerals, iron sulfide, and iron-nickel metal accounting for ~85% of the volume of these meteorites in terms of modal mineralogy (McSween et al. 1991). Figure 12 compares the Didymos MIRI spectrum to spectra taken from the RELAB database of synthetic olivine (specimen ID DD-MDD-086, Fa 10.5 Fo 89.5, particle size < 45 µm) from Dyar et al. (2009), synthetic pyroxene (specimen ID DL-CMP-003-A, Fs25 En75, particle size < 45 µm) from Klima et al. (2007), and a natural diopside (DD-MDD-074, Dyar PI). Klima et al. (2007) and Dyar et al. (2009) synthesized a variety of olivine and pyroxene compositions, and Figure 12 shows the RELAB measurements for the compositions closest to what was measured for Didymos by Dunn et al. (2013) using near-infrared measurements (Fa 24, Fs 19). The laboratory measurements were made in reflectance, and were converted to pseudo-emissivity via Kirchhoff's Law. In addition, because these were measured in ambient conditions, a shift of 0.08 µm to shorter wavelengths was applied to the olivine and pyroxene spectra and a shift of 0.09 µm to shorter wavelengths was applied to the diopside spectrum in order to simulate observations in vacuum, with the shift amounts based on the shifts seen on fine-grained samples by Bramble et al. (2020). We acknowledge that this shift to shorter wavelengths is more or less equal and opposite to the shift expected in CF position by space weathering of these materials (Glotch et al. 2015, section 6.2.1). Iron sulfide is featureless in these wavelengths, and iron metal is both featureless and has a low emissivity, and so neither mineral is included in Figure 12.

The features in the Didymos emissivity spectrum are consistent with the features seen in the laboratory mineral spectra, and do not seem to require minerals other than olivine, low-Ca pyroxene, and high-Ca pyroxene. Without performing a full intimate mineral mixture analysis, we suggest that these 3 minerals can account for all the spectral features observed in the Didymos spectrum. The feature near 10.5 µm mentioned in Section 6.2.1 and the broadened CF, absent from the OC spectra, both appear consistent with an increased contribution to the spectrum from olivine as compared to OC meteorites. Given the complex nature of thermal emission, it is not obvious whether this increased olivine contribution is best interpreted as a compositional or physical effect.

We note the turnup in emissivity shortward of 7 μm. While the calibrations for MIRI are ongoing (Section 4.2), there is no evidence in its extensive characterization campaign suggesting this emissivity rise is an artifact or miscalibration (Wright et al 2023). The contribution of reflected light at the shortest MIRI wavelengths is < 3%, and is accounted for in the thermal modeling. Telescopic data from airless surfaces that cover the 5-7 μm region are rarely found in the literature, but several of the asteroids in the Marchis et al. (2012) sample of Spitzer spectra (Section 6.2.3) have similar emissivity rises at these wavelengths, as does the Infrared Space Observatory (ISO) spectrum of 10 Hygiea (Barucci et al. 2002) and a spectrum of 1 Ceres from the Stratospheric Observatory for Infrared Astronomy (SOFIA) (Vernazza et al. 2017).

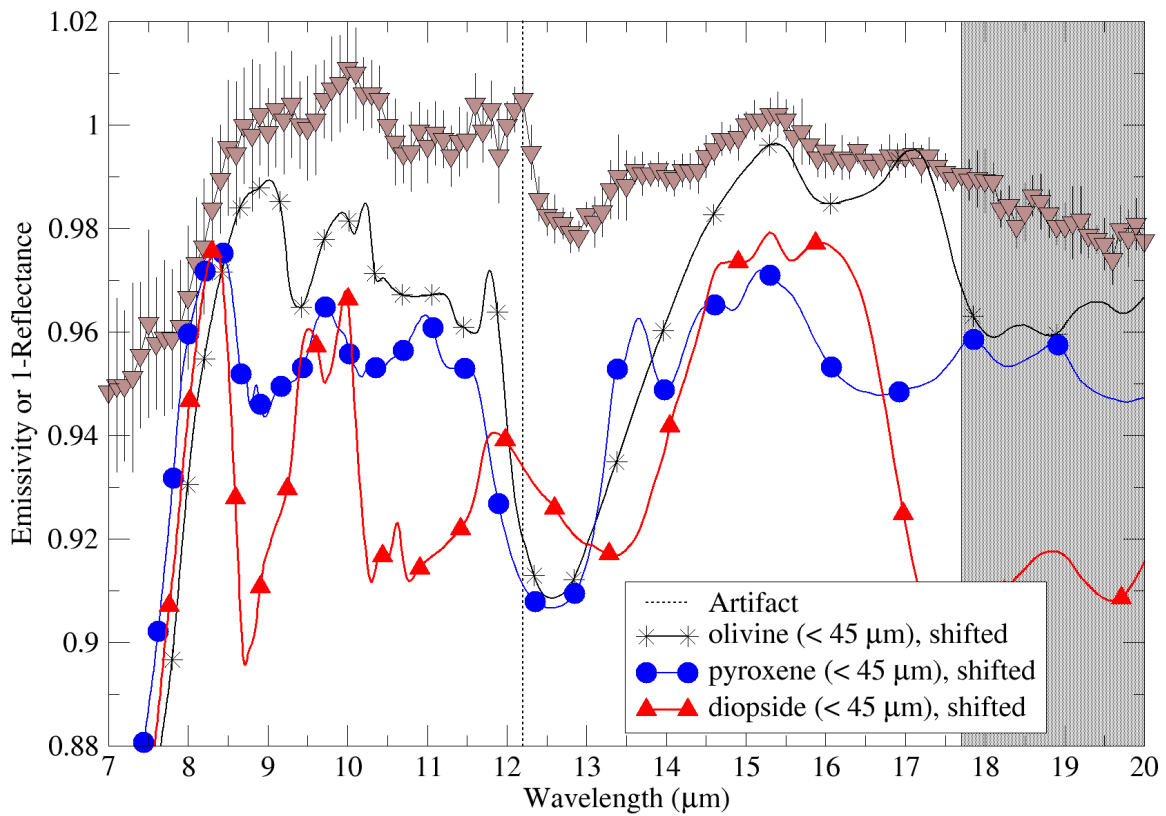

Figure 12: 7—20-μm spectra of Didymos and OC-relevant minerals. The minerals were taken from the RELAB database and measured in reflectance, and (1-reflectance) is plotted for them. The most prominent features in Didymos' emissivity spectrum fall at wavelengths where features can be found in olivine, (low-Ca) pyroxene, or diopside. Features near 10 and 17.5 μm seen in Didymos' spectrum but not the OC spectra in Figure 11 are consistent with olivine.

The best interpretation of this emissivity rise, if real, is not obvious. Silicates have absorption bands (and thus emissivity peaks) in this wavelength region, but they are at longer wavelengths than what is seen here. The spectral consequences of nanophase iron, present in space-weathered materials, have not yet been directly studied at ~5-6 μm, although the effects of darkening have been studied by using nanophase carbon as a darkening agent at slightly longer wavelengths (Shirley et al. 2023). The influence of thermal gradients and small-scale roughness, especially in the situation found on Didymos where thermal skin depths are similar to particle size, may also play a role. Additional laboratory and observational work, as well as additional characterization of the MIRI instrument, will be necessary to understand the nature of this potential emissivity peak in telescopic asteroid spectra, but such studies are beyond the scope of this work.

*6.2.3 Comparison to asteroids*

The mid-infrared spectra of the S-complex asteroid population varies more widely than their spectra in the 0.5—2.5-µm region, which has been interpreted by Vernazza et al (2010) to be due to differences in particle size and/or a space weathering effect. Marchis et al. (2012) studied the mid-IR spectra of asteroid multiple systems, including 10 systems that were either S or Sq types. They found variability between S-complex asteroids (which in their work included 3 V-type asteroids) in terms of their emissivity spectra, which they interpreted as implying diversity in composition and surface properties. Nevertheless, they found all 13 "S-complex" asteroids in their sample to have an emissivity minimum at 7 µm (as did the C-complex asteroids in their sample), and low-contrast transparency features between 11.8—14.2 µm. The features found in the MIRI spectra of Didymos (see Figure 9) are consistent with the Marchis et al. (2012) S-complex dataset.

Figure 13 compares the MIRI spectrum of Didymos to Spitzer spectra of (7) Iris and (433) Eros from Vernazza et al. (2010) and the binary (5407) 1992 AX system from Marchis et al. (2012). Two of these were included because their compositions are thought to be similar to Didymos': Eros has been interpreted as L/LL chondrite using elemental data from the NEAR Shoemaker spacecraft augmented by infrared spectra (Trombka et al. 2001, McCoy et al. 2001), while Iris is a large S-class asteroid, with an LL composition thought most probable by Noonan et al. (2022). Sanchez et al. (2013) interpret (5407) 1992 AX as an S-class asteroid with a pyroxene-dominated basaltic achondrite composition based on its 0.4–2.5-µm reflectance spectrum, and is included as representative of S-class binary NEAs. All are similar to Didymos in most overlapping wavelength regions, demonstrating that Didymos' mid-IR spectrum is consistent with what has already been reported for S-complex asteroids.

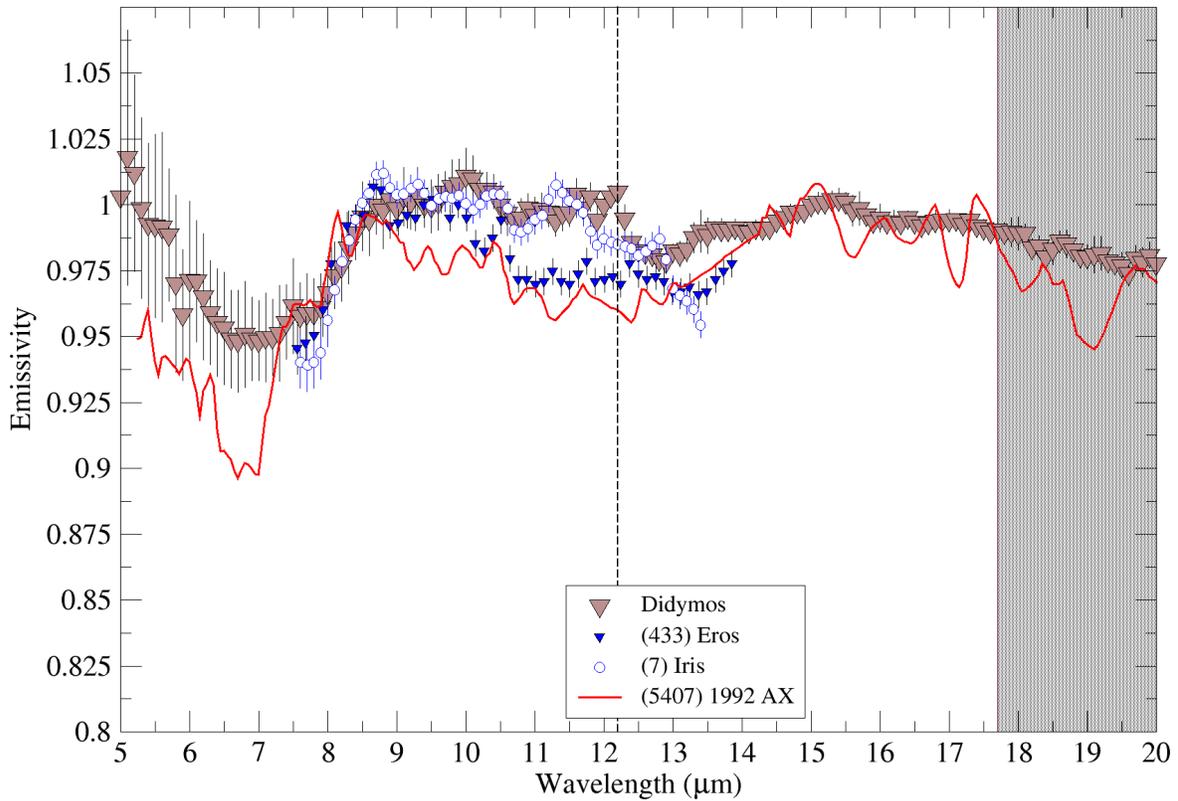

Figure 13: The Didymos spectrum compared to the literature spectra of (433) Eros, (7) Iris (Vernazza et al. 2010) and (5407) 1992 AX (Marchis et al. 2012). Didymos' behavior is at least qualitatively similar to these asteroids, and has spectral behavior within the range seen in this population. The spectra of Didymos, Eros, and Iris are normalized to equal 1 near 9.3 µm, that of 1992 AX is normalized to match Didymos' spectrum near 14.5 µm.

While interpreting the mid-infrared spectroscopy requires fine (<25 µm) particles to explain the presence of a TF, the thermal inertia seems to indicate mm-sized grains (Section 5). Fine-grained material has a smaller thermal inertia and gets hotter than coarse-grained material, and thus the peak temperatures on Didymos will be found in fine-grained material and fine-grained material will provide more thermal flux than coarse-grained material on a per-mass basis (Edgett and Christensen 1991, Harris and Lagerros 2002). Because of this, there may be a bias in the spectra toward fine-grained material that doesn't represent the relative abundance of fine-grained vs. coarse-grained material on Didymos' surface. This apparent mismatch may also be reminiscent of what was found on Bennu by Hamilton et al. (2021): spectral evidence for fine particles while the thermal inertia is higher than one would expect for a fine particulate surface. Turning to Dimorphos rather than Didymos for possible insights, we note that small particles aren't seen on Dimorphos' surface in DART imagery (Daly et al. 2023). However, Dimorphos appeared as smooth as Didymos when imaged at the relatively coarse resolution available for the latter body (Daly et al. 2023). We also note that JWST was primarily observing Didymos' equatorial region, which has a smooth-appearing ridge.

6.3     Summary of Compositional Interpretations

The NIRSpec and MIRI data both point to Didymos' composition being consistent with ordinary chondrites. The comparisons we have shown do not all point to the same OC group, but the majority point toward an LL composition. To be more specific, the BAR is most consistent with LL chondrites (Dunn et al. 2010), the CF position is also most consistent with LL chondrites (Salisbury et al. 1991), the position of the 15-µm emissivity peak is consistent with LL chondrites and less so with H and L chondrites (Bramble 2021), and there is at least a qualitative match in the 0.6—2.5-µm spectrum of Didymos and the recent LL chondrite fall Chelyabinsk. Previous 0.5—2.5-µm measurements of Didymos also have been interpreted as indicating an L/LL composition (Dunn et al. 2013). On the other hand, the position of the Band 1 center for Didymos (and Chelyabinsk) is more suggestive of an H or L chondrite composition.  As noted above, the main mismatches in direct comparison of features in mid-IR OC emissivity spectra to those in the emissivity spectrum of Didymos can be attributed to a higher contribution from olivine in Didymos' spectrum, and olivine is certainly expected to be present based on the BAR value and S-class taxonomic assignment.  Whether that apparent additional olivine is a true compositional difference from OC materials or a result of physical factors is unclear, but the consistency of the BAR value and inferred ol/(ol+px) suggests the latter. The lack of any hydration features near 3 µm, while not diagnostic, is also consistent with our expectations for an OC composition. As an S-class object rather than a Q-class object, we can expect Didymos to have experienced space weathering and have nanophase iron affecting its spectrum, but the shift in CF seen in space-weathered lunar materials is small enough compared to the observational uncertainty that we cannot determine whether it is seen in the MIRI spectrum.

# 7. Summary and Implications for Hera

One of the features of the Didymos system from a planetary defense standpoint is its very common composition–ordinary chondrites account for ~80% of total meteorite falls (Graham et al. 1985, Burbine et al., 2002), which allows the DART results to be taken as representative of a large fraction of potential impactors. The spectroscopic characterization of Didymos from JWST reinforces the interpretation of the system as typical of NEOs. Didymos' spectrum is consistent with an anhydrous surface, as expected for an ordinary chondritic composition, and its thermal inertia is typical for other NEOs of its size. In addition, the rock-forming minerals of ordinary chondrites are well known and exhibit well characterized mechanical properties that are consistent with similar asteroids (Moyano-Cambero et al., 2017; Tanbakouei et al., 2019). Its mid-infrared spectrum is within the range of what is seen in the literature for S-complex asteroids, and it is qualitatively similar to ordinary chondrite powders (and LL chondrite powders specifically) across most of the 0.6—20-µm spectral range.

The Hera mission is scheduled to arrive at the Didymos system in late 2026 and can follow up or extend several of the results presented here. Hera specifically will be able to image Didymos' surface at higher spatial resolution (Michel et al. 2022) than was possible from DART and LICIACube (Dotto et al. 2023, in revision), providing a test of the regolith particle size inferences derived from mid-infrared spectroscopy and a check on the measured thermal inertia.

Additional observations with JWST would also be useful. For instance, it is possible that a very weak absorption is present near 3 µm, obscured by observational uncertainties. Higher-quality data in that spectral region would allow a tighter constraint on Didymos' hydration state or detect an absorption band, which would in turn give insight into space weathering or parent body processes on Didymos (and presumably other S-complex NEOs). Observations could be timed so that MIRI and NIRSpec were observing the same part of Didymos, and/or could be designed to try to take maximum advantage of mutual events to extract information about Dimorphos specifically.

However, there are additional questions that will likely require not just Hera's upcoming rendezvous or additional observations but advances in laboratory work and spectral modeling. For instance, why there are mismatches in detail between Didymos and other S-complex NEOs on the one hand and their presumed-analog meteorites and constituent minerals on the other hand is still an open question. We look forward to Hera and future JWST measurements of additional S-complex asteroids to help us continue efforts to understand the population of potential asteroid impactors, for the science return and to help inform planetary defense efforts to mitigate potential collisions.

## 8. Acknowledgements

This work was supported by the DART mission, NASA Contract No. 80MSFC20D0004. This work is based on observations made with the NASA/ESA/CSA JWST. The data were obtained from the Mikulski Archive for Space Telescopes at the Space Telescope Science Institute, which is operated by the Association of Universities for Research in Astronomy, Inc., under NASA contract NAS 5-03127 for JWST. These observations are associated with program 1245. ASR and CAT would like to especially thank the people at the Space Telescope Science Institute who worked tirelessly and enthusiastically to enable the program of JWST observations of Didymos to be carried out. IW's research was supported by an appointment to the NASA Postdoctoral Program at the NASA Goddard Space Flight Center, administered by Oak Ridge Associated Universities under contract with NASA. SNM and HBH acknowledge support from NASA JWST Interdisciplinary Scientist grant 21-SMDSS21-0013. BR acknowledges funding support from the UK Science and Technology Facilities Council (STFC). JMTR acknowledges financial support from project PID2021-128062NB-I00 funded by Spanish MCIN/AEI/10.13039/501100011033. AL, MP, JRB, GP acknowledge financial support from Agenzia Spaziale Italiana (ASI-INAF contract No. 2019-31-HH.0 and No. 2022-8-HH.0). The JWST spectra presented in this paper were obtained from the Mikulski Archive for Space Telescopes (MAST) at the Space Telescope Science Institute. The specific observations analyzed can be accessed via https://doi.org/10.17909/9ymc-em60.